\documentclass{optica-article}
\journal{opticajournal} 
\articletype{Research Article}
\usepackage{url}

\usepackage{lineno}

\begin{document}

\title{Observation of  electric field induced superradiance slowdown in ultracold Rydberg atomic gases}

\author{Yunhui He,\authormark{1} Jingxu Bai,\authormark{1,2} Yuechun Jiao,\authormark{1,2} Weibin Li,\authormark{3,4} and Jianming Zhao\authormark{1,2,*}}

\address{\authormark{1}State Key Laboratory of Quantum Optics and Quantum Optics Devices, Institute of Laser Spectroscopy, Shanxi University, Taiyuan 030006, China\\
\authormark{2}Collaborative Innovation Center of Extreme Optics, Shanxi University, Taiyuan 030006, China\\
\authormark{3}School of Physics and Astronomy, and Centre for the Mathematics and Theoretical Physics of Quantum Non-equilibrium Systems, University of Nottingham, Nottingham, NG7 2RD, UK\\
\authormark{4}weibin.li@nottingham.ac.uk}

\email{\authormark{*}zhaojm@sxu.edu.cn} 

\begin{abstract*} 

Atoms excited to electronically high-lying Rydberg states decay to low-energy states through spontaneous emission processes. We investigate the impact of a static electric field on the superradiant emission process between Rydberg $|60D_{5/2}\rangle$ and $|61P_{3/2}\rangle$ states in an ultracold Cesium Rydberg atom ensemble. We report experimental observations of a significant slowdown in superradiance upon applying an electric field. To understand the slowing down of superradiance, we employ a master equation in which Stark effects and collective decay are taken into account. We solve the master equation with the discrete truncated Wigner approximation (DTWA) method. Our numerical simulations demonstrate that superradiance decoherence is caused by the Stark shifts of the Rydberg level. Our theoretical simulations qualitatively match the experimental observations. Our work provides new insights into controlling quantum critical behaviors, with implications for quantum many-body dynamics, and the study of quantum phase transitions.

\end{abstract*}

\section{Introduction}
Rydberg atoms, with their electrons orbiting far from the nucleus in highly excited states, exhibit exaggerated properties like large electric dipole moments and strong interactions~\cite{gallagher1994rydberg}. These characteristics make them ideal candidates for studying quantum phenomena and place them at the forefront of quantum physics research. The study of Rydberg atoms opens up new possibilities for advancements in quantum information processing~\cite{lukin2001dipole,saffman2010quantuma,paredes-barato2014alloptical,shao2024rydberg}, quantum simulations~\cite{weimer2010rydberg,scholl2021quantuma,scholl2023erasure}, single-photon devices~\cite{baur2014singlephotonb,gorniaczyk2016enhancement}, and precision measurements~\cite{kaubruegger2019variational,hu2022continuously}. Among the many fascinating applications of Rydberg atoms, their role in the exploration of superradiance stands out. 

Superradiance is a phenomenon where dense atoms collectively emit coherent photons at an enhanced rate due to cooperative interactions~\cite{dicke1954coherence,lehmberg1970radiation,gross1982superradiance,ficek2002entangled}, which are beneficial for the development of narrow linewidth lasers~\cite{bohnet2012steadystate,norcia2016coldstrontium}, quantum metrology~\cite{wang2014heisenberg,liao2015gravitational} and atomic clocks~\cite{norcia2016superradiance}. Therefore superradiance has been studied in various systems experimentally and theoretically, such as Rydberg atoms~\cite{gross1979maser,wang2007superradiance,hao2021observation}, trapped ions~\cite{devoe1996observation}, Bose-Einstein condensates~\cite{inouye1999superradiant,lode2017fragmented}, cavity~\cite{mlynek2014observation,suarez2022superradiance} and arrays of quantum emitters~\cite{masson2022universality}. 
Rydberg atom systems provide a unique intrinsic advantage for studying superradiance because the wavelength $\lambda$ of transition between Rydberg energy levels is on the order of millimeters, which is significantly larger than the typical interatomic separation $R\sim\mu\text{m}$, i.e., satisfying the Dicke limit condition of $R\ll\lambda$~\cite{dicke1954coherence}. That makes Rydberg atoms an ideal platform for investigating superradiant behavior~\cite{hao2021observation,sutherland2017superradiance,suarez2022superradiance}. 
In the cold Rydberg atomic system, van der Waals (vdW) interactions  tend to modify 
the superradiant dynamics and its scaling~\cite{hao2021observation}, 
whereas dipole-dipole interactions can suppress superradiance due to dephasing effects~\cite{sutherland2017superradiance,suarez2022superradiance}. The competition between interactions and dissipation not only affects superradiant emission but also provides a unique lens through which to study time crystal~\cite{wu2024dissipative} and quantum phase transitions~\cite{baumann2010dicke,zhu2020squeezed,he2022superradianceinduced,ferioli2023nonequilibrium}. Superradiance reveals critical phenomena related to the collective behavior of many-body quantum systems, such as transitions between ordered and disordered states. In our previous work, we have observed blackbody radiation (BBR) enhanced superradiance in ultracold Rydberg gases and directly measured the temporal evolution of superradiant decay between Rydberg states $|{nD}\rangle$ and $|{(n + 1)P}\rangle$ in free space~\cite{hao2021observation}. However, the effect of external fields, such as electric fields, introduces additional complexity and remains underexplored sufficiently. 

In this work, we investigate the effect of a static electric field on the superradiance in the Rydberg atomic system. Our observations reveal that an electric field leads to a slowdown in superradiant emission. We attribute this effect to the Stark shift, which alters the energy levels of the Rydberg atoms and thus modifies their collective emission properties. To gain deeper insights into these observations, we have employed a theoretical model to simulate the impact of electric fields on superradiance. Our findings provide a valuable understanding of how external fields interact with superradiance in Rydberg systems and offer new perspectives on controlling the temporal evolution of quantum states and understanding how external fields can be used to modify behaviors of Rydberg many-body systems.

\begin{figure}[htpb]
\centering  
\includegraphics[width=0.9\linewidth]{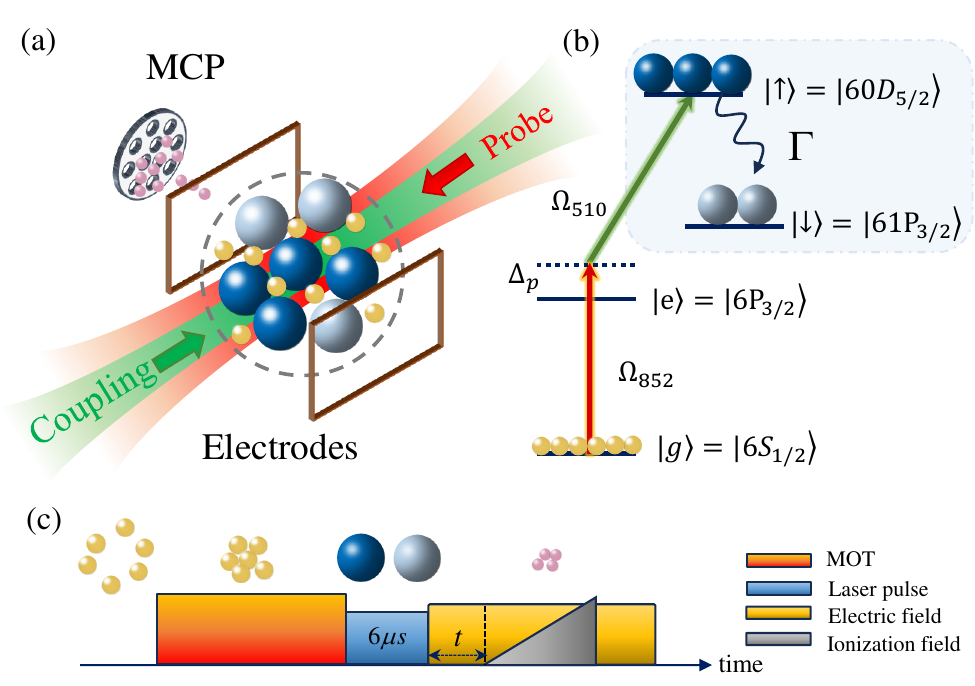}	
\caption{(a) Sketch of the experimental setup.  Rydberg state $nD_{5/2}$ of Cesium atoms is prepared in a spherical magneto-optical trap (MOT) (not shown). The first-step ($\lambda_p=852~\rm{nm}$) and the second-step ($\lambda_c=510~\rm{nm}$) excitation lasers counter-propagate through the MOT center performing two-photon Rydberg excitation. A pair of electrodes are used to apply a static electric field for interacting with the Rydberg atom and apply a ramp field for the state selective field ionization of Rydberg state. The Rydberg dynamics are detected by multichannel plate (MCP) after field ionization. (b) Energy level diagram. The first 852-nm laser, $\Omega_{852}$, drives the transition from the ground state $|{g}\rangle=|{6S_{1/2},F=4}\rangle$ to the intermediate excited state $|{e}\rangle=|{6P_{3/2},F^\prime=5}\rangle$ with  blue detuning $\Delta_p=2\pi\times360$~MHz, while the second 510-nm laser, $\Omega_{510}$, couples the transition of $|{e}\rangle$ $\to$ $|{\uparrow}\rangle=|{60D_{5/2}}\rangle$. The atoms in state $|{\uparrow}\rangle$ decays to a neighboring Rydberg state $|{\downarrow}\rangle=|{61P_{3/2}}\rangle$ with decay rate $\Gamma$. (c) Experimental timing sequence. We change the interaction time $t$ to investigate the evolution dynamics of the Rydberg state. }
\label{model}
\end{figure}

\section{Experimental observation of superradiance with electric field}

In our experiment,	$^{133}$Cs ground-state atoms ($N$ up to $10^7$) are laser-cooled to $\sim100~\mu\rm{K}$ and trapped in a spherical magneto-optical trap (MOT). The atomic cloud has a diameter of $\sim550~\mu\rm{m}$, which is much smaller than the transition wavelength $\lambda=92.93~\rm{mm}$ of $|60D_{5/2}\rangle\to|61P_{3/2}\rangle$~\cite{hao2021observation}. Two laser fields, probe laser (wavelength $\lambda_p=852$~nm) and coupling laser ($\lambda_c=510$~nm), are counter-propagated through the MOT center [see Fig.~\ref{model}(a)]. The laser fields, with respective  Rabi frequency $\Omega_p=2\pi\times59.1$~MHz and $\Omega_c=2\pi\times6.4$~MHz, drive two-step transitions between the ground state $|{g}\rangle=|{6S_{1/2},F=4}\rangle$ and the Rydberg state $|{\uparrow}\rangle=|{60D_{5/2}}\rangle$ via the intermediate excited state $|{e}\rangle=|{6P_{3/2}, F^{\prime} = 5}\rangle$. Due to the effect of spontaneous emission and blackbody emission, the atoms would decay to the neighboring Rydberg state $|{\downarrow}\rangle$ = $|61P_{3/2}\rangle$ with a decay rate $\Gamma$. The energy level scheme is shown in Fig.~\ref{model}(b).  The laser frequencies are stabilized by utilizing a super-stable optical Fabry-Perot (FP) cavity with fineness $\mathcal{F}=15000$, and the laser linewidth is less than 50~kHz. The probe $852$~nm laser is blue shifted $\Delta_p=2\pi\times360$~MHz from the intermediate level $|e\rangle$ using a double-pass AOM. The probe and control lasers with the linear polarization have a $1/e^2$ beam waist of $\omega_p=80~\mu$m and $\omega_c=40~\mu$m, respectively, forming a cylindrical excitation region. Three pairs of electrodes encircle the excitation region (only one pair of electrodes is shown in Fig.~\ref{model}), allowing us to compensate the stray electric fields using Stark spectroscopy. The CST modelling the electric field in the MOT chamber shows that the electric field in the MOT center by applying the potential at the electrodes is uniform.

The timing sequence is shown in Fig.~\ref{model}(c). After switching off the MOT beams, we turn on the excitation laser pulse for $6~\mu$s to excite ground atoms to $|60D_{5/2}\rangle$ Rydberg state. Then we apply a weak static electric field for interaction time $t$ with Rydberg atoms, and finally turn on a ramp electric field with the ramp time 3~$\mu$s for the state selective field ionization of the Rydberg atoms, which enables to distinguish the laser excited  $|60D_{5/2}\rangle$ state and decayed $|61P_{3/2}\rangle$ state. Prior to measuring the superradiant evolution of Rydberg atoms, we first calibrate the MCP ion detection system using shadow images taken before and after laser excitation. By comparing these images, we determine the number of Rydberg excitations and get the gain factor of our ions detection system. On the other hand, we further apply a 3.2-GHz microwave field to couple the $|60D_{5/2}\rangle \to|61P_{3/2}\rangle$ transition and use a microwave spectroscopy to verify the superradiance mainly appears in the $|60D_{5/2}\rangle$ to $|61P_{3/2}\rangle$ decay channel. More experimental details can be found in Ref.~\cite{hao2021observation}.  

\begin{figure}[htpb]
\centering  
\includegraphics[width=1.0\linewidth]{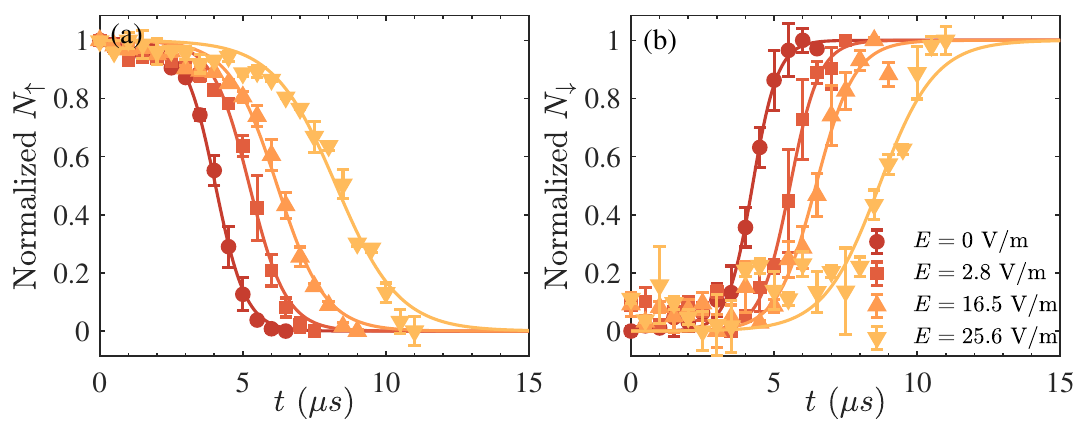}	
\caption{Evolution of normalized Rydberg atom population (a) $N_\uparrow$ and (b) $N_\downarrow$ with the indicated static electric field for the initial $|60D_{5/2}\rangle$ atom number $N_e=33400$ (the formula $y=(y-y_{min})/(y_{max}-y_{min})$ is used to normalize experimental data). For the case of without electric field (dark color),  the atom population $N_\uparrow~(N_\downarrow)$ decreases (increases) slowly for interaction time $t<3~\mu s$, and drastically evolves when  $3~\mu s<t<6~\mu s$, driven by superradiance~\cite{gross1982superradiance}.  When $t \geq 6~\mu s$, the atom population approaches to the minimum (maximum) value. For the cases with the static field, the evolution dynamics process becomes slower with increasing the electric field. The experiment data (dots) and the fitting curves obtained by Eq.~(\ref{eq:N_down_alpha}) (solid lines) agree well.}
\label{Dynamics_experiment}
\end{figure}

To investigate the population evolution of Rydberg atoms, we change the interaction time $t$ after Rydberg excitation pulse and detect the time-resolved ions signal of $N_\uparrow$ and $N_\downarrow$ states. Furthermore, to obtain the evolution dependence on the electric field, during the interaction time, we apply a weak static electric field. In Fig.~\ref{Dynamics_experiment}, we present the normalized population of $N_\uparrow$ in Fig.~\ref{Dynamics_experiment}(a)  and $N_\downarrow$  in  Fig.~\ref{Dynamics_experiment}(b) for indicated electric fields and the initial $|60D_{5/2}\rangle$ atom number $N_e=33400$. In the absence of an electric field, $E=0$~V/m, both the populations of $\lvert\uparrow\rangle$ state and $\lvert\downarrow\rangle$ state undergo a fast dynamics process, which includes three distinct regions. In the first region, $0<t<3~\mu s$, the population varies at a very slow rate, indicating that the collective emission of Rydberg atoms is initially weak. In the second region, $3<t<6~\mu$s, the population displays a pronounced change in the dynamics, with the population $N_{\uparrow}$ ($N_\downarrow$) experiencing a rapid decline (increase), which corresponds to the onset of strong superradiant emission as the atoms coherently emit photons in a burst. When $t \geq 6~\mu s$, the population of $N_{\uparrow}$ remains almost constant, suggesting that the superradiant burst has ended where further emission is minimal~\cite{hao2021observation}. It is noted that the population of $N_{\downarrow}$ would decrease after reaching the maximum value as atoms in this state would further decay to lower states due to their limited lifetime. As our experiment is only focused on the population evolution of superradiance between the two Rydberg states, we do not show further decay processes of the $N_{\uparrow}$ and $N_{\downarrow}$ in the  Fig.~\ref{Dynamics_experiment}.  

It is found, from Fig.~\ref{Dynamics_experiment}, that the population decay process slows down when a weak field is applied, and the population dynamic displays much slower with increasing the electric field. When $E = 25.6~\text{V/m}$, the system takes $t \approx 13~\mu\text{s}$ to reach a constant, which is nearly twice that in the absence of an electric field. The measurements in the Fig.~\ref{Dynamics_experiment} demonstrate that the superradiance of $|nD_{5/2}\rangle\to |(n+1)P_{3/2}\rangle$ transition is strongly suppressed by the external electric field. 

\section{Model and theory}

In order to explain experimental observations of Fig.~\ref{Dynamics_experiment}, we consider a two-level model, consisting of levels $|{\uparrow}\rangle$ and $\lvert\downarrow\rangle$ with the decay rate $\Gamma$, shown with the shallow region in Fig.~\ref{model}(b). The decay dynamics of the system are governed by the following master equation~\cite{ficek2002entangled}
\begin{equation}\label{eq:master1}
\dot{\rho}(t)=\mathcal{L}[\rho(t)],
\end{equation}
where $\rho$ is the many-body density matrix, and operator $\mathcal{L}(\rho)$ describes the dissipation, 
\begin{equation}\label{eq:lindblad}
	\mathcal{L}(\rho)=\sum_{j,k}^N\Gamma_{jk}\left[\hat{\sigma}_-^j\rho\hat{\sigma}_+^k-\frac{1}{2}\left\{\hat{\sigma}_+^k\hat{\sigma}_-^j,\rho\right\}\right],
\end{equation}
where $\hat{\sigma}_{+}^j=\lvert{\uparrow_j}\rangle\langle_j\downarrow\rvert\left(\hat{\sigma}_{-}^j=\lvert{\downarrow_j}\rangle\langle_j\uparrow\rvert\right)$ is the raising (lowering) operator for the site $j$ and $\Gamma_{jk}$ is the decay rate. The single-atom decay rate includes spontaneous radiation and blackbody radiation (BBR), normally coupling neighboring Rydberg states. This decay rate is given approximately by $\Gamma_{j}\approx \Gamma_{0}+\Gamma_{BBR}$, where the $\Gamma_{0}$ is the spontaneous decay rate in vacuum, defined as $\omega^3_j\mu^2_j /3\pi\epsilon_0\hbar c^3$ with $\omega_j$ the transition frequency and $\mu_j$ the dipole moment~\cite{ficek2002entangled}. The calculated transition of $|60D_{5/2}\rangle \to |61P_{3/2}\rangle$~\cite{daniel2024cesium} is $\Gamma_j \approx 62$~Hz. In addition, due to the average spacing between Rydberg atoms much smaller than $\lambda$, the spatial dependence has negligible effect, thus $\Gamma_{jk}=\Gamma_{j}=\Gamma$.

\begin{figure}[htpb]
\centering  
\includegraphics[width=1\linewidth]{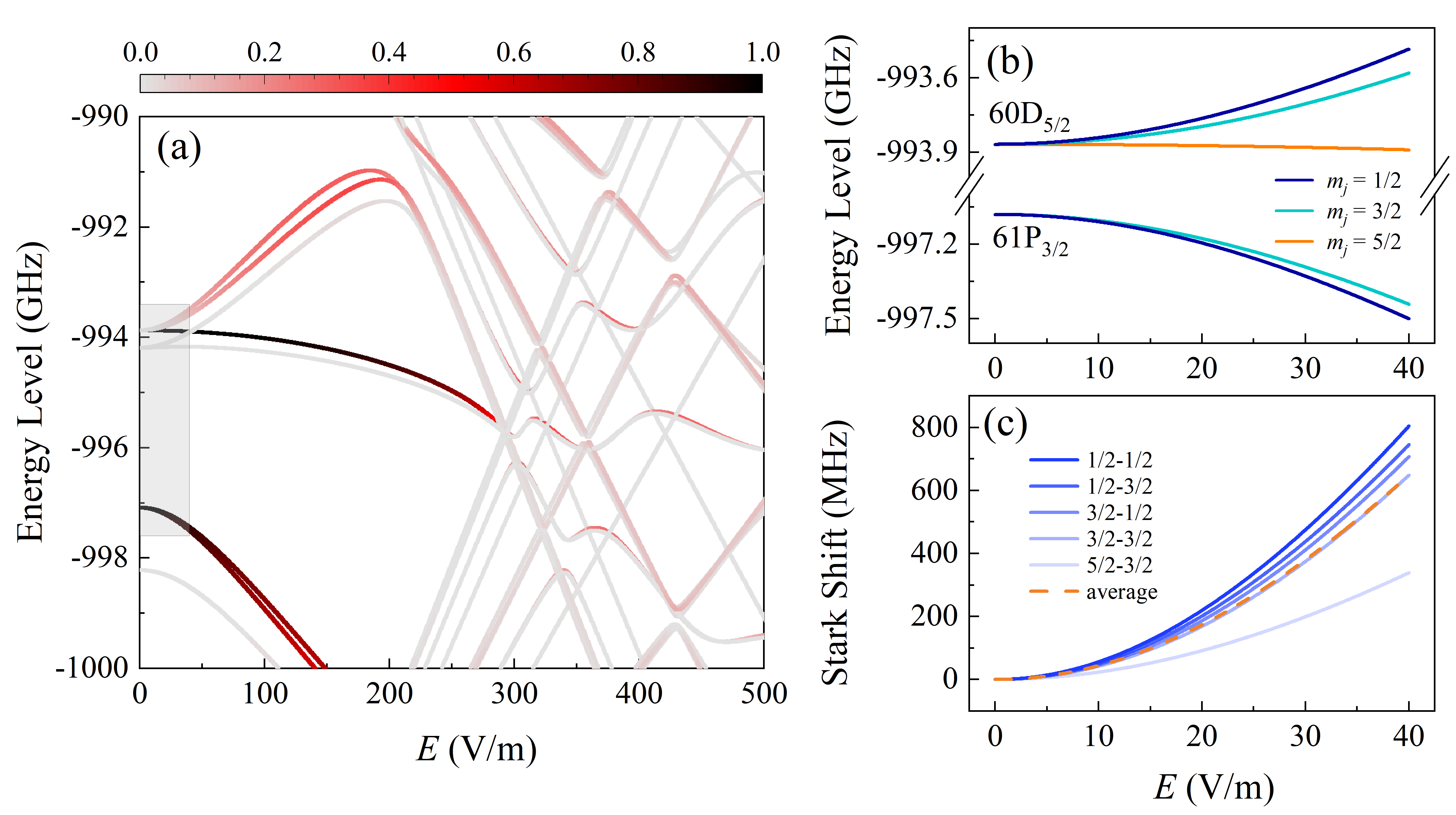}	
\caption{(a) The calculated Stark map  for states $|{60D_{J}}\rangle$ and $|{61P_{J}}\rangle$ in the range of $E < 500~\rm{V/m}$. To show the effect of state mixing, the levels are colored proportional to the fraction of the target state present in each eigenstate. (b) Zoom-in of the gray shadow of (a) for $60D_{5/2}$ and $61P_{3/2}$ states,  different colors represent different $m_j$ values. (c) System Stark shifts as a function of the electric field $E$. The solid lines represent the shift of different magnetic energy levels and the dashed line represents the average value of Stark shift for five different transitions.}
\label{starkshift}
\end{figure}

In the presence of a weak external electric field, Rydberg states experience the Stark shift. We first calculate the Stark map by using the Alkali Rydberg Calculator (ARC) package~\cite{robertson2021arc}. The Stark map for states $|{60D_{J}}\rangle$ and $|{61P_{J}}\rangle$ is shown in Fig.~\ref{starkshift}(a). When the electric field is weak, $E < 200$~V/m, we primarily observe energy level splittings and shifts, with the most significant proportion indicated by color. As the electric field increases, $|{60D_{J}}\rangle$ and $|{61P_{J}}\rangle$ levels mix with nearby high-$l$ states.  The Stark map shows complex patterns, accompanied by energy level crossings and avoided crossings. This complexity arises primarily because higher $l$-states are more sensitive to the external field. The transition to high-$l$ state has a relatively smaller proportion. In our experiment, we do not observe excitation of these states. Here, we focus on the shaded region in Fig.~\ref{starkshift}(a). We extract the most prominent Stark shifts of $|{60D_{5/2}}\rangle$ and $|{61P_{3/2}}\rangle$ states, which are displayed in Fig.~\ref{starkshift}(b), with different colors representing different $m_j$ values. Given our interest in the frequency shift difference between these two energy levels, we further calculated the Stark shifts for the five possible transitions according to the selection rules and plotted in Fig.~\ref{starkshift}(c) to illustrate the electric field-induced shifts more clearly. Since it is not feasible to separate these transitions in our experiment, we consider average  Stark shifts in the two-level model. The average of five possible Stark shifts is represented by the dashed line in Fig.~3(c).

With the Stark shifts at hand, they contribute to an effective detuning $\Delta$ to our system. The master equation (\ref{eq:master1}) can be rewritten as ($\hbar=1$)
\begin{equation}\label{eq:master2}
\dot{\rho}(t)=-i\left[\sum_{k=1}^N\frac{\Delta}{2}\hat{\sigma}_z^k,\rho(t)\right]+\mathcal{L}[\rho(t)],%
\end{equation}
where $\hat{\sigma}_z=|{\downarrow}\rangle\langle{\downarrow}|-|{\uparrow}\rangle\langle{\uparrow}|$. For a many-body system ($N\gg 1$), it is difficult to solve the quantum master equation by direct diagonalization. We use mean-field (MF) approximation to decouple the many-body density matrix $\rho\approx \Pi_i\rho_i$ by neglecting quantum correlations between different sites~\cite{diehl2010dynamical}. The MF equations of motion of atoms read, 
\begin{subequations}\label{eq:mean-field}
\begin{align}
\dot{\mathbf{s}_x^k}&=-\mathbf{s}_z^k\sum_{j=1}^N\Gamma_{jk}\mathbf{s}_x^j-\Delta \mathbf{s}_y^k,\label{eq:MF-a}\\
\dot{\mathbf{s}_y^k}&=-\mathbf{s}_z^k\sum_{j=1}^N\Gamma_{jk}\mathbf{s}_y^j+\Delta \mathbf{s}_x^k,\label{eq:MF-b}\\
\dot{\mathbf{s}_z^k}&=\sum_{j=1}^N\Gamma_{jk}(\mathbf{s}_x^j\mathbf{s}_x^k+\mathbf{s}_y^j\mathbf{s}_y^k),\label{eq:MF-c}
\end{align}
\end{subequations}
where $\mathbf{s}_\eta^k=\langle S_\eta^k\rangle$ is the expectation value of spin operator $S_\eta^k(\eta=x,y,z)$. We then apply the discrete truncated Wigner approximation (DTWA) to describe quantum many-body dynamics by introducing quantum fluctuations into the initial states~\cite{schachenmayer2015manybody,schachenmayer2015dynamics}. 

To obtain the dynamical evolution of Rydberg population, we consider all atoms initially in the upper state $\lvert\uparrow\rangle$ with the density matrix $\hat{\rho}_{0}^k = \lvert\uparrow\rangle\langle\uparrow\rvert$, resulting in a fixed classical spin component along $z$ ($\sigma_{z}^k = -1/2$) and fluctuating spin components in the $x$ and $y$ directions ($\sigma_{x(y)}^k \in \{-1/2, 1/2\}$, each with $50\%$ probability). Mean values of observables, such as the Rydberg population, are calculated by averaging over many trajectories. In the simulation, an ensemble of Rydberg atoms separated by the blockade radius $R_b$ with a Gaussian spatial distribution is considered. The number of trajectories $N_{traj}=5000$ are used to achieve convergence of the DTWA simulations.

\begin{figure}[htpb]
\centering  
\includegraphics[width=1.0\linewidth]{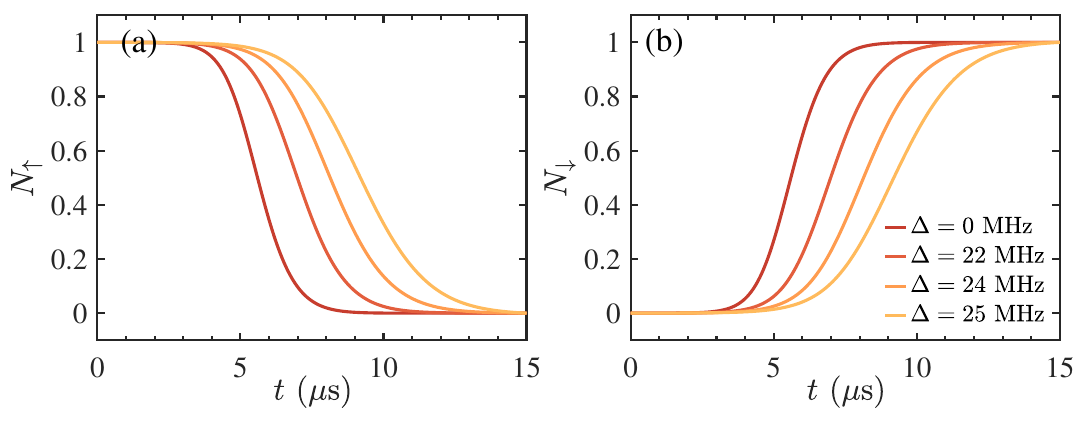}	
\caption{The calculated dynamical evolution of Rydberg population (a) $N_{\uparrow}$ and (b) $N_{\downarrow}$ with the detuning, $\Delta$, different Stark shift with the DTWA method. Other parameter: $N_t = 33400, N_{traj} = 5000$. }
\label{Dynamics_theory}
\end{figure}

We numerically solve Eqs.~(\ref{eq:mean-field}) based on the DTWA method with $N_t=33400$ (Rydberg number in the calculation). The dynamics of the Rydberg population for indicated detunings, $\Delta$, are shown in Fig.~\ref{Dynamics_theory}. It is seen that the dynamics evolve most rapidly for $\Delta=0$, e.g. in the absence of an electric field. As the detuning increases, the population evolution process gradually slows down, which means superradiance is suppressed due to electric field induced Stark shifts. The theoretical simulation qualitatively agrees with the experimental measurements in Fig.~\ref{Dynamics_experiment}.

\section{Results and discussions}

Rydberg population can be obtained analytically from the master equation~(\ref{eq:master1}) when there are no external fields~\cite{gross1982superradiance},
\begin{equation}\label{eq:N_down_alpha}
N_{\downarrow}=\frac{N_t}{2}+\frac{N_t}{2}\tanh{\left[\frac{\Gamma_{col}(t-t_d)}{2}\right]},
\end{equation}
where $\Gamma_{col}=N_t\Gamma$ is the collective decay rate of atomic system, and $t_d=\ln(N_t)/[\Gamma(N_t+1)]$ is the delay time. By turning on the electric field, the dynamics will be modified. As the electric field is relatively weak,  we fit the experimental data with Eq.~(\ref{eq:N_down_alpha}), in which $\Gamma_{col}$ and $t_d$ are free parameters. These two parameters characterize the dependence of the superradiance on the Stark effects.  

\begin{figure}[htpb]
\centering  
\includegraphics[width=1.0\linewidth]{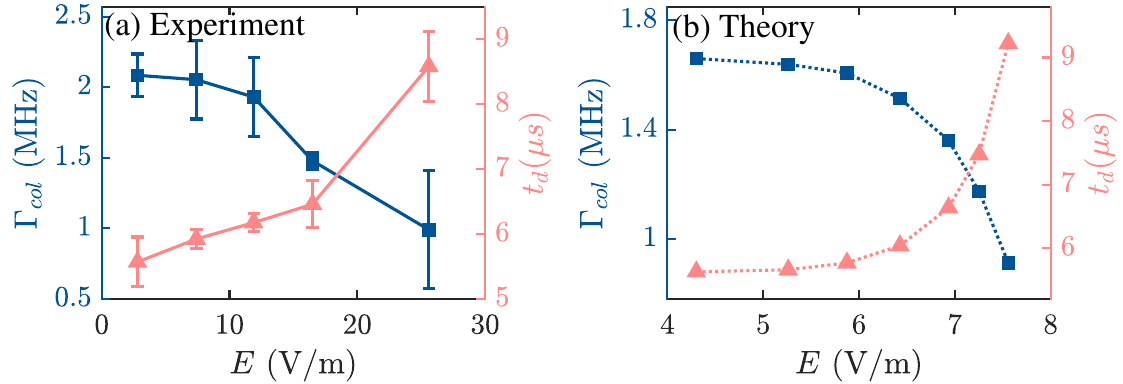}	
\caption{The collective decay rate $\Gamma_{col}$ and delay time $t_d$, obtained by the fittings of Eq.~(\ref{eq:N_down_alpha}) to the experimental measurements of Fig.~\ref{Dynamics_experiment} in (a) and theoretical simulations of Fig.~\ref{Dynamics_theory} in (b). The fitted $\Gamma_{col}$ and $t_{d}$ change monotonically with the electric field $E$ both for measured and calculated data. 
}
\label{fit_compare}
\end{figure}

In Fig.~\ref{fit_compare}(a) we show the fitting parameters, $\Gamma_{col}$ and $t_d$, extracted from the fitting to experimental data shown in Fig.~\ref{Dynamics_experiment} with the electric field $E$, where the left Y-axis represents the fitted collective decay rate $\Gamma_{col}$ and the right Y-axis represents the fitted superradiance lifetime $t_d$. In the absence of an electric field, we obtain $\Gamma_{col}\approx2.2~$MHz where $N_t$ is approximately found to be $35600$ in the fitting, consistent with the value used in our calculations. When we add the external static electric field, the $\Gamma_{col}$ declines, showing the decrease with the electric field increases, which means that the coupling between the Rydberg atoms and electric field significantly restrains the superradiance process. Meanwhile, the decay time $t_d$ of the atomic system increases rapidly with the electric field.  

To compare with theory, in Fig.~\ref{fit_compare}(b), we present $\Gamma_{col}$ and $t_d$ from the fitting to the simulated dynamic process in Fig.~\ref{Dynamics_theory}. From the Stark spectrum calculated in Fig.~\ref{starkshift}(c), we can obtain the scaling of detuning $\Delta$ and electric field strength $E$. It is seen that when electric field strength $E$ increases, $\Gamma_{col}$ ($t_d$) demonstrates a monotonic decrease (increase) which demonstrates a similar dependence with the experiment results. The difference between the theoretical calculation and experimental data could be caused by two factors. Firstly, the electric field value is read by the output of a voltage source divided by the space of the two grids, which may have a deviation from the field sensed by the atoms, as the grids have a hole in the center for the trapping beam going through. Secondly, the automatic ionization of Rydberg atoms can also create an extra electric field, which is difficult to account for. Overall, from Fig.~\ref{fit_compare}, our theoretical calculations qualitatively agree with the experimental results.

It should be noted that vdW interaction and dipole-dipole interaction always play a crucial role when considering Rydberg states. In our previous work~\cite{hao2021observation}, we investigated the BBR enhanced superradiance, where we accounted for vdW interactions, which were found to change the superradiant dynamics and modify the scaling of the superradiance. In the current analysis, we have omitted the vdW interaction primarily because the energy level shifts due to the vdW interaction are much less than the Stark shift due to the external electric field that is on the order of $100$~MHz, which significantly exceeds the magnitude of the interaction effects. On the other hand, the presence of dipole-dipole interactions would slow down the superradiance due to many-body dephasing and even may lead to a complete collapse of superradiance~\cite{gross1982superradiance,suarez2022superradiance}. Given our aim to use a simplified model to elucidate the impact of the electric field on superradiance, we have chosen to neglect collective interaction terms
in this context, 
and our model captures the characteristic trends observed in the experiments. The competition of these two effects can be explored by, e.g., changing the density of the atomic gas, or considering weaker electric fields. This is worth exploring in the future. 

\section{Conclusion}

In conclusion, our experimental observations have revealed a significant slowdown in superradiant emission when an electric field is applied to ultracold Rydberg atomic gases. Using the DTWA method, we have calculated the mean-field master equation and performed theoretical simulations. Our results qualitatively show that the Stark effect, induced by the electric field, leads to a shift in the energy levels of the Rydberg atoms, thereby slowing down the superradiant dynamics. The qualitative agreement between our theoretical simulations and experimental data confirms the validity of our model and enhances our understanding of how external fields can influence collective emission properties in Rydberg systems. This study provides a new insight into controlling the temporal evolution of Rydberg states and paves a way for future research on quantum phase transitions with Rydberg atom ensembles in different electromagnetic fields.

\begin{backmatter}
\bmsection{Funding}
National Natural Science Foundation of China (U2341211, 62175136, 12241408, and 12120101004); Innovation Program for Quantum Science and Technology (2023ZD0300902); Fundamental Research Program of Shanxi Province (202303021224007); the 1331 project of Shanxi Province. W.L. acknowledge the Engineering and Physical Sciences Research Council (EP/W015641/1) and the Going Global Partnerships Programme of the British Council (IND/CONT/G/22-23/26).

\bmsection{Disclosures}
The authors declare no conflicts of interest.

\bmsection{Data availability} Data underlying the results presented in this paper are not publicly available at this time but may be obtained from the authors upon reasonable request.
\end{backmatter}

\bibliography{main}

\end{document}